\documentstyle[epsfig]{mn}
\begin{document}
\LARGE
\normalsize

\title[Radio emission from Cygnus~X-1]
{Orbital modulation and longer-term variability in the radio emission
from Cygnus~X-1}

\author[Pooley, Fender \& Brocksopp]
{G.G. Pooley,$^1$\thanks{Email ggp1@cam.ac.uk} R.P. Fender,$^2$ C. Brocksopp$^3$\\
$^1$ Mullard Radio Astronomy Observatory, Cavendish Laboratory,
Madingley Road, Cambridge CB3 0HE\\
$^2$ Astronomical Institute `Anton Pannekoek' and Center for High Energy
Astrophysics, University of Amsterdam, Kruislaan 403, \\
1098 SJ Amsterdam, The Netherlands. \\
$^3$ Astronomy Centre, CPES, University of Sussex, Falmer, Brighton,
BN1 9QJ\\}

\maketitle

\begin{abstract}
20 months of observations of the radio emission at 15 GHz from Cygnus~X-1,
starting in
1996 October, show variations at the binary period of 5.6 days, but with a
phase offset from those at X-ray wavelengths. There are also longer-term
variations on a time-scale of 150 days which are only loosely related to the
soft X-ray flux. The source was in the hard/low X-ray state throughout this
period. The mean 15-GHz flux density is 13 mJy, the radio spectrum is flat, and
the semi-amplitude of the orbital modulation about 2 mJy. We discuss the
possible origins of the modulation and the relationship to the soft X-ray
emission.

\end{abstract}

\begin{keywords}
radio continuum: stars -- X-rays: stars -- stars: individual: Cygnus X-1
\end{keywords}

\section{Introduction}

Cygnus~X-1 (= HDE 226868, V1357 Cyg) is a well-studied X-ray binary;
Lewin, van Paradijs \& van den Heuvel (1995)
give a recent summary of the data on this system.
The compact object is massive, $M \geq 7 M_{\sun}$, and has long been
considered as one of the most likely candidates for an identified black hole.
The donor is a
supergiant with $M \geq 20 M_{\sun}$. The binary period, 5\fd 6, is
well-established from optical observations (e.g. Gies \& Bolton 1982)
and has also been found in soft and hard X-rays
(2--12 keV: Zhang, Robinson \& Cui 1996;
20--100 keV: Paciesas et al. 1997).

The X-ray emission has a complex spectrum and varies between `hard' and
`soft' states, often referred to (in terms of their soft X-ray flux) as
`low' and `high' respectively.
The radio observations described here were taken
exclusively during the more usual hard/low state. X-ray variations are
found on all time-scales from milliseconds to years.
For the initial comparison with soft X-ray data, in this paper we use
the \it{RXTE }\rm all-sky monitor `quick-look' database, covering 2--12 keV.

Radio emission from Cyg~X-1 has been studied since soon after its discovery
as an X-ray source. In 1971, the flux density increased from an undetectable
level to a fairly steady state at the same time as a change in the
X-ray state (Hjellming \& Wade 1971; Braes \& Miley 1971).
Generally, the radio emission has
a flat spectrum and a flux of 10--20 mJy when the source is in the hard/low
X-ray state and is lower when in the soft/high state.
A radio event associated
with an X-ray flare in 1975 was reported by Hjellming, Gibson \& Owen
(1975). Han (1993) and Hjellming \& Han (1995) report modulation at a
period which they believed to be the orbital period (and which was
confirmed with the early part of the present data by Fender, Brocksopp
\& Pooley 1997a).

\section{Observations}

The Ryle Telescope at the Mullard Radio Astronomy Observatory,
Cambridge was used for the monitoring reported here. Details are as
for the observations reported by Pooley \& Fender (1997); in this case
the phase calibrator used was B2005+403.
The radio data are presented here as 10-min integrations
and the typical rms uncertainty is
1~mJy.  The data shown here were collected between 1996 Oct 21 and
1998 Jun 13.  We have also made use of radio data from the
Green Bank Interferometer (details of which may be found in
Waltman et al. 1994) which monitors this and many other
sources at 2.25 and 8.3 GHz.

\section{Results}

\begin{figure*}
\leavevmode\epsfig{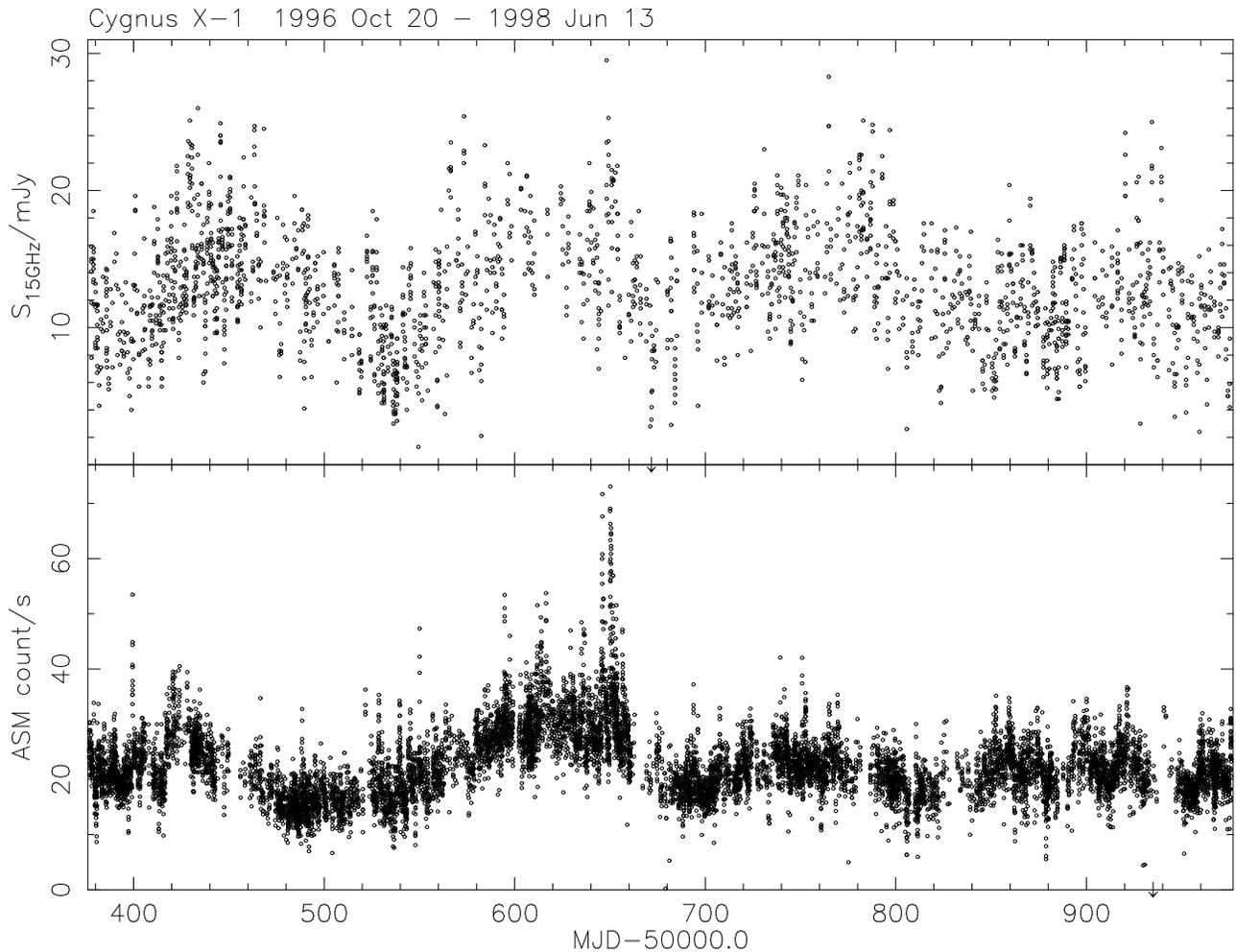}
\caption{20 months' of 15-GHz radio flux-density measurements (top)
and 2--12 keV X-ray monitoring (bottom) of
Cyg~X-1.}
\end{figure*}

\begin{figure*}
\leavevmode\epsfig{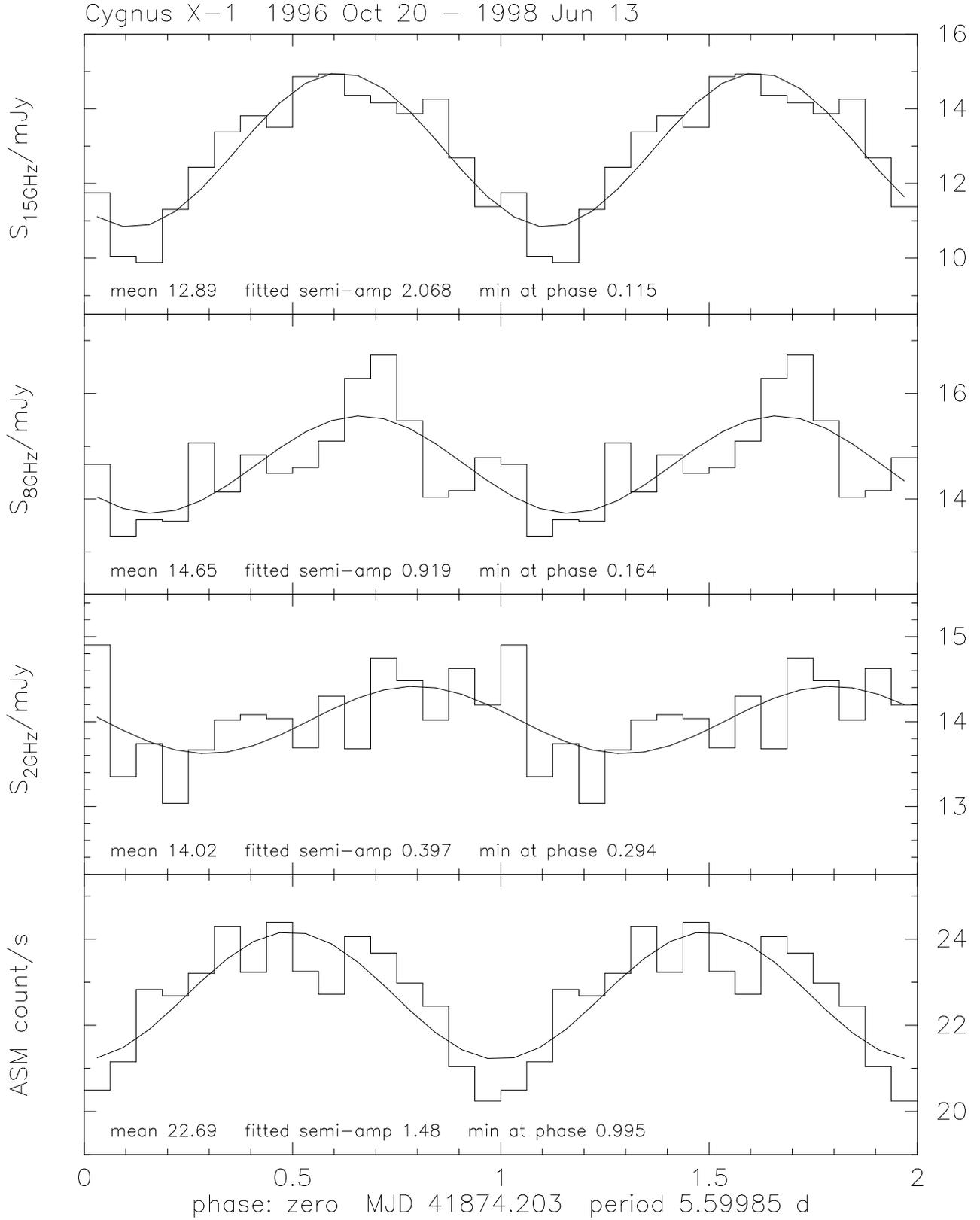}
\caption{Radio flux densities at 15, 8.3 and 2.25 GHz,
together with XTE ASM 2--12 keV data,
folded into 16 equal phase bins using the ephemeris of Tarasov
(see text) with phase zero defined as inferior conjunction of the
supergiant. Overdrawn
on each plot are best-fit sinusoids. At radio wavelengths the
fractional modulation decreases towards lower frequencies. A phase lag
is also evident in the radio data, increasing towards lower
frequencies.}
\end{figure*}

The whole dataset is plotted in Fig. 1, together with the \it{RXTE }\rm ASM data
for the same period.  The X-ray emission was in the hard/low state for the
whole of this interval. Activity in the previous few months took it
into the high/soft state, with considerable increase and rapid variations
in the mean soft X-ray flux
(Zhang et al. 1997); unfortunately, our monitoring program had
not started and the Green Bank Interferometer was not operating during
the period of increased X-ray activity.

In displaying the 5\fd 6 period, we have used the spectroscopic ephemeris
recently derived by A.E.\ Tarasov (Brocksopp et al. 1998):


\[
     P = 5\fd 599847 \pm 0\fd 000018
\]
\[
     T = {\rm JDh}\  2441874.703 \pm 0.009
\]
and the folded plots all refer to this epoch, inferior conjunction of the
supergiant, as the zero of phase.
Other recent discussions of the ephemeris (e.g. Sowers et al. 1998,
LaSala et al. 1998) are in agreement with these values.
In Fig. 2 we show four sets of data folded on this ephemeris
into 16 equal phase bins: these are the radio flux densities at 15, 8.3
and 2.25~GHz, and the \it{RXTE }\rm ASM 2--12~keV count-rate.

There are several features which are apparent from these folded plots:

\begin{itemize}

\item{The overall mean radio spectrum is flat, as has been noted before.}

\item{The modulation at the 5\fd 6 period is stronger at the higher
frequencies, and is barely detectable at 2.25 GHz. At 15 GHz, the
semi-amplitude is 16\% of the mean; in the ASM band it is 6.5\%.}

\item{The minimum of the X-ray flux agrees very closely with the spectroscopic
determination of the conjunction.}

\item{The minimum of the radio emission is significantly later than that of the
X-rays (by about 16 h at 15 GHz, and longer delays at lower frequencies).}

\item{There is a suggestion that the folded 15-GHz data do not fit particularly
well to a sinusoid.  Residuals after subtracting the best-fitting
sinusoid are shown in Fig. 3. The deviations appear only marginally
significant; there are 2165 points in the 15-GHz database, and so
typically 135 per bin in these plots. The rms scatter throughout the
database relative to the mean is 4.2~mJy, and reduces to 4.0~mJy on
subtraction of the sinusoid.  So the expected deviation of the points
in the residual, {\it if} the bulk of the variations are independent of the
5\fd 6 periodicity, would be about 0.35~mJy. The rms deviation is in fact
0.6~mJy, suggesting that there may be an additional non-sinusoidal
term.}
\end{itemize}

By far the most striking feature in the longer-term variation is the
apparent periodicity at about 150 -- 160 d, obvious on
inspection of Fig. 1, which faded away after the first two or three
cycles.
We note that this `period' is approximately half the  294~d
X-ray and optical period reported by Priedhorsky, Terrel \& Holt
(1983) and Kemp et al. (1987).   We only have 600 days of radio data, but we
find no evidence for a 294-day period.

The variations in radio flux density do seem to be loosely correlated
with the X-ray flux (Fig. 1).
Fig. 4 shows a plot of daily averages of the two datasets
(all data-points within one calendar day are averaged, and 415 of the
600 days have data from both wavebands).
A plot of the same datasets after subtraction of the best-fit sinusoids
has a very similar appearance; the 5\fd 6 period is not responsible
for the correlation in Fig. 4. There is also one isolated event
in Fig. 1 which shows evidence for a relationship between the two wavebands;
the largest flux densities in both the 15-GHz and ASM datasets are near
MJD 50650 (1997 July 21). Fig. 5 is an expanded plot of this section of data,
where it is apparent that typical variations in both wavebands are on
short timescales and that essentially continuous monitoring
would be required for any more detailed interpretation.

\begin{figure}
\leavevmode\epsfig{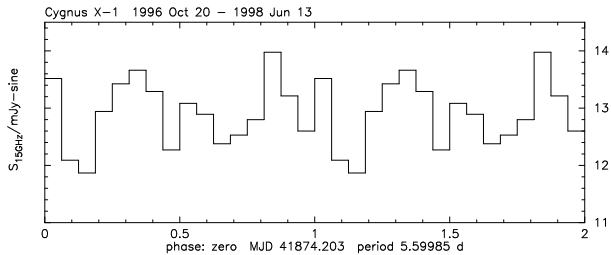}
\caption{Residuals of 15 GHz data folded as in Fig. 2 but with best-fit
sinusoid removed. The scatter of the residuals suggests that the radio
modulation may not be consistent with a simple sinusoid alone.}
\end{figure}

\begin{figure}
\leavevmode\epsfig{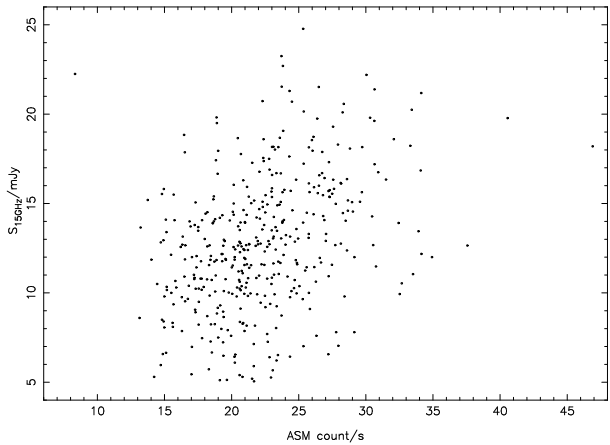}
\caption{Daily averages of the 15-GHz and ASM data.}
\end{figure}

\begin{figure}
\leavevmode\epsfig{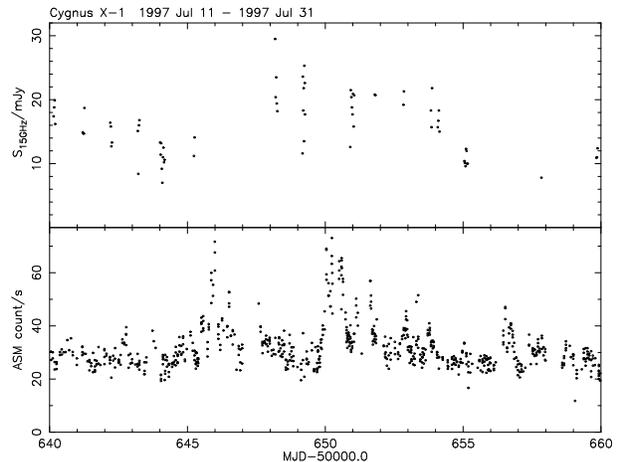}
\caption{Details of the 15-GHz and ASM data near MJD 50650, including
the highest intensities recorded during the 600-day interval.
The typical rms noise on the radio flux densities is 1~mJy.}
\end{figure}

\section{Discussion}


Hjellming \& Han (1995) discuss models for the the radio emission of X-ray
binaries. They comment on the difficulty of producing satisfactory models
for stable, flat-spectrum emission, and that optically-thick, conical jets
do represent a possible class of model. A relativistic jet whose orientation
was linked to the underlying binary system would provide a natural mechanism
for the 5\fd 6 period. It is less clear how that model would explain the change
in phase and amplitude of the 5\fd6 variation with the observing frequency.
If the periodic variations are not associated with a relativistic beam,
we may consider changes
in the absorbtion along the line of sight (including eclipses).
In the case of eclipses, the lower modulation at longer wavelengths may reflect
a larger volume contributing to the emission as a result of some optical-depth
effect.

The X-ray emission of Cyg~X-1, in particular the state transitions,
has many similarities with that of another black hole candidate X-ray
binary, GX~339-4 (Tanaka \& Lewin 1995 and references therein).
GX~339-4 has recently been found to
exhibit radio emission with a flat spectrum and comparable radio
luminosity to Cyg~X-1 (Sood \& Campbell-Wilson 1994; Fender et
al. 1997b). This radio emission appears to be correlated with XTE ASM
monitoring during low/hard states in a similar manner to that of Cyg
X-1 (Hannikainen et al. 1998) and furthermore has been observed to
decrease by more than a factor 20 during a transition from hard/low to
soft/high state (Fender et al., in prep.). The radio emission from
GX~339-4 is also roughly correlated with the hard X-ray (20--100 keV)
during the low/hard state (Hannikainen et al. 1998) and we might
expect to see such a correlation in Cyg~X-1 also.

Observations of these sources in both radio and X-ray regimes may well shed
more light on the interrelation between the disc/corona and the
synchrotron (presumably outflowing) regions of the binaries.

Further analysis of the Cygnus~X-1 observations,
including data from more wavebands, will be described
in another paper.

\section*{Acknowledgements}

RPF is an EC Marie Curie Fellow supported by grant ERBFMBICT 972436.
The Ryle Telescope is supported by PPARC.
We acknowledge with thanks the use of the quick-look results provided by
the ASM/RXTE team, and Green Bank Interferometer data.
The Green Bank Interferometer is a facility of the National
Science Foundation operated by the NRAO in support of
NASA High Energy Astrophysics programs.  CB acknowledges a PPARC studentship.

\end{document}